\documentclass[conference]{IEEEtran}
\IEEEoverridecommandlockouts

\usepackage[utf8]{inputenc} \usepackage{adjustbox}
\usepackage{algorithmic}
\usepackage{amsmath, amssymb,amsfonts}
\usepackage{array}
\usepackage{booktabs}
\usepackage{caption}
\usepackage{cite}
\usepackage{colortbl}
\usepackage{graphicx}
\usepackage{nomencl}
\usepackage{siunitx}
\usepackage{subcaption}
\usepackage{todonotes}
\usepackage{texnames}
\usepackage{textcomp}
\usepackage{xcolor}
\usepackage{xurl}
\usepackage[colorlinks=true, urlcolor=blue, linkcolor=black]{hyperref}
\usepackage{orcidlink}

\graphicspath{{./}}
\begin{document}

\title{Geometric Shape Modelling and Volume Estimation of Dry Bulk Cargo Piles using a Single Image
\thanks{Norwegian Research Council contract number 326609}
}

\author{
\IEEEauthorblockN{Debanshu Ratha\orcidlink{0000-0003-4377-8915}}
\IEEEauthorblockA{\textit{Environmental Impacts Section}\\\textit{Akvaplan-niva AS}\\
            Tromsø, Norway\\
            der@akvaplan.niva.no\\
            \url{https://orcid.org/0000-0003-4377-8915}
        }
\and
\IEEEauthorblockN{Madhu Koirala\orcidlink{0009-0005-8872-5392}}
\IEEEauthorblockA{\textit{Department of Electrical Engineering}\\
\textit{UiT The Arctic University of Norway, Campus Narvik}\\
Narvik, Norway\\
m.koirala@uit.no\\
\url{https://orcid.org/0009-0005-8872-5392}
        }
\and[\hfill\mbox{}\par\mbox{}\hfill]
\IEEEauthorblockN{Pål Gunnar Ellingsen\orcidlink{0000-0002-3331-5581}}
\IEEEauthorblockA{\textit{Department of Electrical Engineering}\\
\textit{UiT The Arctic University of Norway, Campus Narvik}\\
Narvik, Norway \\ 
pal.g.ellingsen@uit.no\\
\url{https://orcid.org/0000-0002-3331-5581}
        }
    }

\maketitle

\begin{abstract}
Volume estimation of onshore cargo piles is of economic importance for shipping and mining companies as well as public authorities for real-time planning of logistics, business intelligence, transport services by land or sea and governmental oversight. In remote sensing literature, the volume of pile is estimated by relying on the illumination property of object to construct the geometric shape from a single image, alternatively, stereographic imaging for construction of a digital elevation model from pairs of images. In a fresh perspective, we propose a novel approach for estimating volume from a single optical image in this work where we use the material property, which relates the base dimensions of the pile to its height through the critical angle of repose. In materials literature, often this is well-studied for fixed base and their \textit{in situ} volume estimation for different materials. In this work, however, we mathematically model the geometric shape of the pile through a fixed height model. This is appropriate because the unloading crane arm that forms the pile can rise only up to a certain height and generally moved in the horizontal plane during unloading of the material. After mathematically modelling the geometric shape of regular piles for fixed heights under rectilinear motion of unloader, we provide closed form formula to estimate their volume. Apart from laying the mathematical foundations, we also test it on real optical remote sensing data of an open bulk cargo storage facility for silica sand and present the results. We obtain an accuracy of $95\%$ in estimating the total bulk storage volume of the storage facility. This is a first demonstration study and will be integrated with applied machine learning approaches or current state-of-art approaches in the future for more complex scenarios for estimating dry bulk cargo pile volume. 
\end{abstract}\hfill\\
\makenomenclature
\nomenclature{\(L_0\)}{Outer rectangle length of the pile at zero height}
\nomenclature{\(W_0\)}{Outer rectangle width of the pile at zero height}
\nomenclature{\(b_0\)}{Radius of base of the right circular cone formed when the height is fixed}
\nomenclature{\(H\)}{Height of the pile when it is loaded to maximum capacity with respect to its base}
\nomenclature{\(\theta_c\)}{Critical angle of repose for the dry bulk cargo material which the pile is made up of}
\printnomenclature
\section{Introduction}
In international dry bulk trade, estimates of how much ore is produced, stored, and consumed carry significant economic value for the mining and shipping industries and governments. This volume information of cargo material accumulated at mines, in ports and harbors, at construction sites, can be used for route planning, government oversight, and market prediction. These applications can then increase profitability for transport companies and reduce their overall carbon footprint due to more efficient transport, a much needed climate action plan for the future of the earth. From a governmental point of view, this monitoring allows for independent verification of reported production and consumption numbers. With their possibility of monitoring large areas on earth, satellites offer a way of continuous monitoring of bulk storages.

Traditional ways of estimating volumes are using stereo imaging or LiDAR scanning, often recorded in close proximity. We have found that studies on the estimation of the volume of dry bulk cargo piles using satellite imagery are almost non-existent in the current remote sensing literature~\cite{alsayed2023stockpile} with the exception of a few notable works~\cite{dAutume2020Stockpile,Marchesconi2022Interactive,gitau2022spatial,Mari2021Automatic} within the monitoring of stockpile volume.  

In~\cite{dAutume2020Stockpile}, d'Autume et al. propose the use of the linear shape-from-shading (SFS) technique for estimating stock pile volume. It relies on the illumination property of object to construct the geometric shape for the stockpiles by solving a first-order nonlinear partial differential equation, which the authors convert to linear form using several practical and theoretical assumptions on the model. It is different from stereo imaging techniques in that it uses minimal input in the form of a single image. The main challenge for the authors was the validation of the volumes with stereo-imaged techniques because of lack of synchronous acquisition of stereo images to generate a digital elevation model (DEM). The method was demonstrated in a time series of PlanetScope imagery acquired over the region of interest over several months. However, the conditions imposed to solve this equation limited the applicability to other scenarios, although it can still be used for volume stock monitoring. The SFS technique was used for interactive segmentation in high-resolution synthetic aperture radar (SAR) images~\cite{Marchesconi2022Interactive}. Here, the SFS technique is applied to solve for the model with respect to range and azimuth plane and demonstrated on X-band SAR images from CapellaSpace. However, a large bias was observed when estimating the volume on the ground. 

In~\cite{gitau2022spatial}, the volume was estimated by first generating contours from control points with 3D coordinates that were then interpolated to create a Triangular
Irregular Network (TIN), and then rasterized to create the DEM. There, the ambiguity in volume arises from the choice of control points. 

In case of techniques that may require more than one acquisition/image, automatic stockpile volume monitoring using stereo-imaging techniques has been demonstrated in~\cite{Mari2021Automatic}. 
The main idea there lies in using date-wise Rational Polynomial Cameras (RPCs) used for multiview stereo-imaging by means of rotation and error compensation for inaccurate satellite altitude information which are then utilized for accurate Digital Surface Model (DSM) generation of the scene, consequently, improving the estimation of volumes in 3D surface models. The method is demonstrated using multiview stereo image time series of SkySat images, and the authors had access to very accurate ground truth for the region of interest.

In a fresh perspective, we propose a novel approach for estimating volume from a single optical image in this work where we use the material property, which relates the base dimensions of the pile to its height through the critical angle of repose. We mathematically model the geometric shape of the pile through a fixed height model. This is appropriate because the unloading crane arm that forms the pile can only rise up to a certain height and is generally moved in the horizontal plane during unloading of the material. After mathematically modelling the geometric shape of regular piles for fixed heights under rectilinear motion of unloader, we provide closed form formula to estimate their volume. Apart from laying the mathematical foundations, we also test it on real optical remote sensing data of an open bulk cargo storage facility for silica sands for which we present the results.
\section{Methodology}\label{sec:Methodology}

\begin{figure}[h!bt]
    \centering
    \includegraphics[width=0.25\columnwidth]{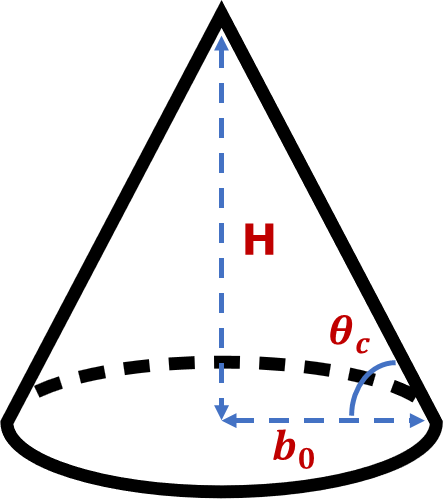}
    \caption{The right circular cone formed by a granular material when released from a single point illustrating the angle of repose denoted as $\theta_c$.}
    \label{fig:conical-pile-angle-of-repose}
\end{figure}

\begin{figure}[h!bt]
    \centering
    \includegraphics[width=0.45\columnwidth]{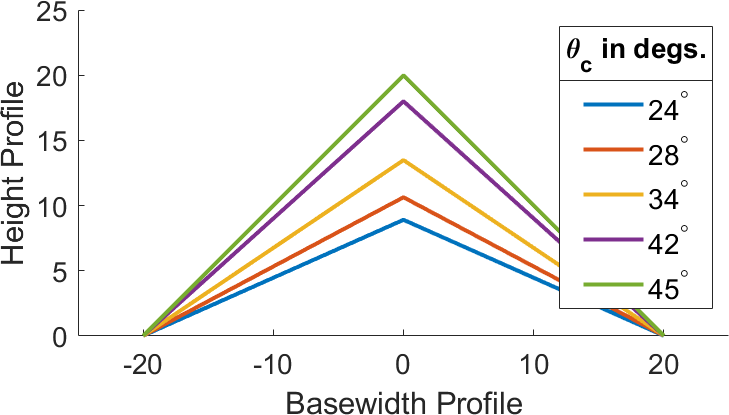}
    \includegraphics[width=0.45\columnwidth]{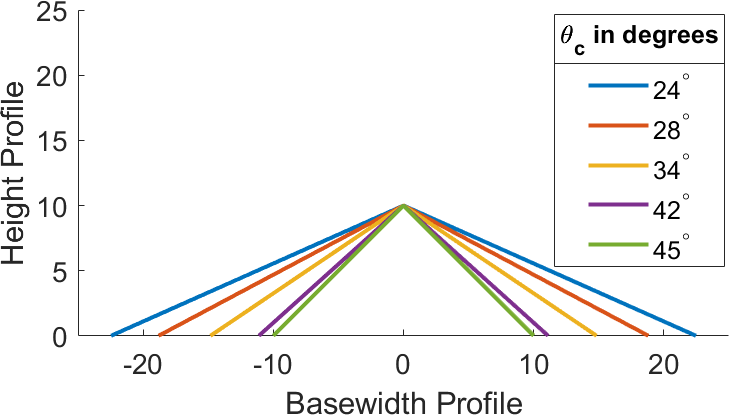}
    \includegraphics[width=0.45\columnwidth]{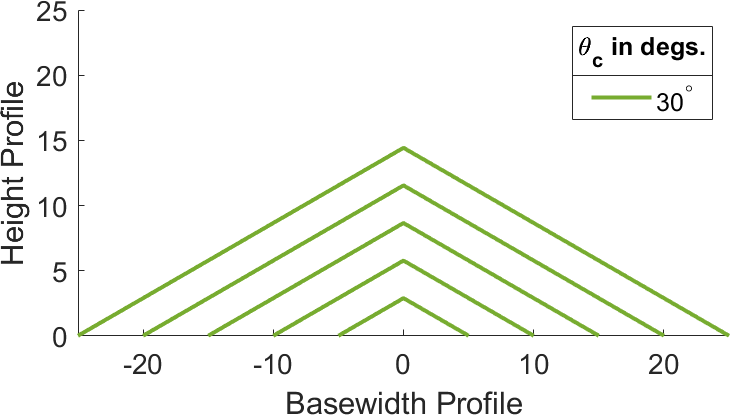}\\\vspace{1pt}
    
    \caption{Cone cross sections for different cases: (clockwise from top to bottom) fixed $\theta_c$, base and height respectively.}
    \label{fig:fix-CS}
\end{figure}

Under normal circumstances, when a granular material is dispensed using a stationary funnel-shaped unloader, it takes the shape of a solid right circular cone as shown in Fig.~\ref{fig:conical-pile-angle-of-repose}. The radius of the circular base $b_0$ is related to the height $H$ of the cone through the angle of repose $\theta_c$. This angle of repose is an intrinsic material property that is dependent on the type of material, its granularity, and environmental parameters like its wetness. If known or estimated \emph{a priori} (we will discuss the implications of this later), it can be leveraged to calculate the volume of piles where only the base (or lateral) dimensions are known, without explicitly knowing the height information.
The well-known relationship between the angle of repose and the height is as follows:
\begin{equation}\label{eq:3par_reln}
\tan{\theta_c} = \frac{H}{b_0}
\end{equation}
Fig.~\ref{fig:fix-CS} shows some of its consequences. In the three cases discussed there, we observe that for the same base, different materials will have different heights because $\theta_c$ varies with the material. The higher the value of $\theta_c$, the higher the height of the pile. In the second image, if the height is fixed, the base width varies for different materials. Here, the smaller $\theta_c$, the wider the base required. And in the last case, for a fixed $\theta_c$ or the same material, the height varies with $b_0$ (base width is twice $b_0$) and vice versa, as determined by equation~\eqref{eq:3par_reln}.

In practice, we will often come across a range of values for $\theta_c$ for a material in literature instead of a single value. This is so because of the finer composition of material, moisture content, packing properties, weight of the pile, etc. Thus, uncertainties in measurement of $b_0$ and $\theta_c$ (which are independent) may propagate to $H$ when indirect estimation is performed. In such a case, the standard deviation in estimation of $H$ is calculated as follows~\cite{kappaupsilon1966notes}:
\begin{equation}
    \sigma_{H} = \sqrt{\left(\sigma_{\theta_c} \times b_0 \times \sec^2{\theta_c}\right)^2 + \left(\sigma_{b_0} \times \tan{\theta_c}\right)^2}
\end{equation}
Assuming we could measure $b_0$ accurately from the image, we have a simplified expression:
\begin{equation}\label{eq:error}
    \sigma_{H} = \sigma_{\theta_c} \times b_0 \times \sec^2{\theta_c}
\end{equation}
Thus, the standard deviation of height depends linearly on the standard deviation of $\theta_c$ with an intercept of zero when base radius is measured accurately. However, the slope of this line depends linearly on $b_0$, but non-linearly on $\theta_c$.

\subsection{Proposed Model}
\emph{In practical settings, however, it may be prudent to fix the height $H$ instead of the base and observe what shape the base takes when loaded to maximum capacity.} This is because the unloader can only be lifted up to a certain height. Additionally, in most cases, it can be moved in the horizontal plane.

\begin{figure}[h!bt]
    \centering
\includegraphics[width=\columnwidth]{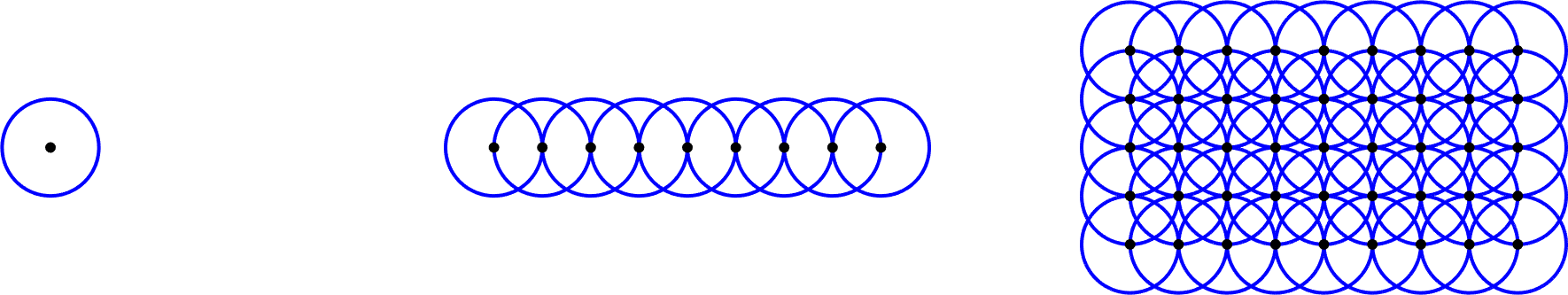}\\
    \caption{The piles visualized as identical overlapping right-circular cones of same height whose peaks lay in the plane of motion forming a grid pattern as above. In limit, the grid separation distance $\rightarrow 0$.}
    \label{fig:grid-patterns}
\end{figure}

\begin{figure}[h!bt]
    \centering     \includegraphics[width=0.32\columnwidth]{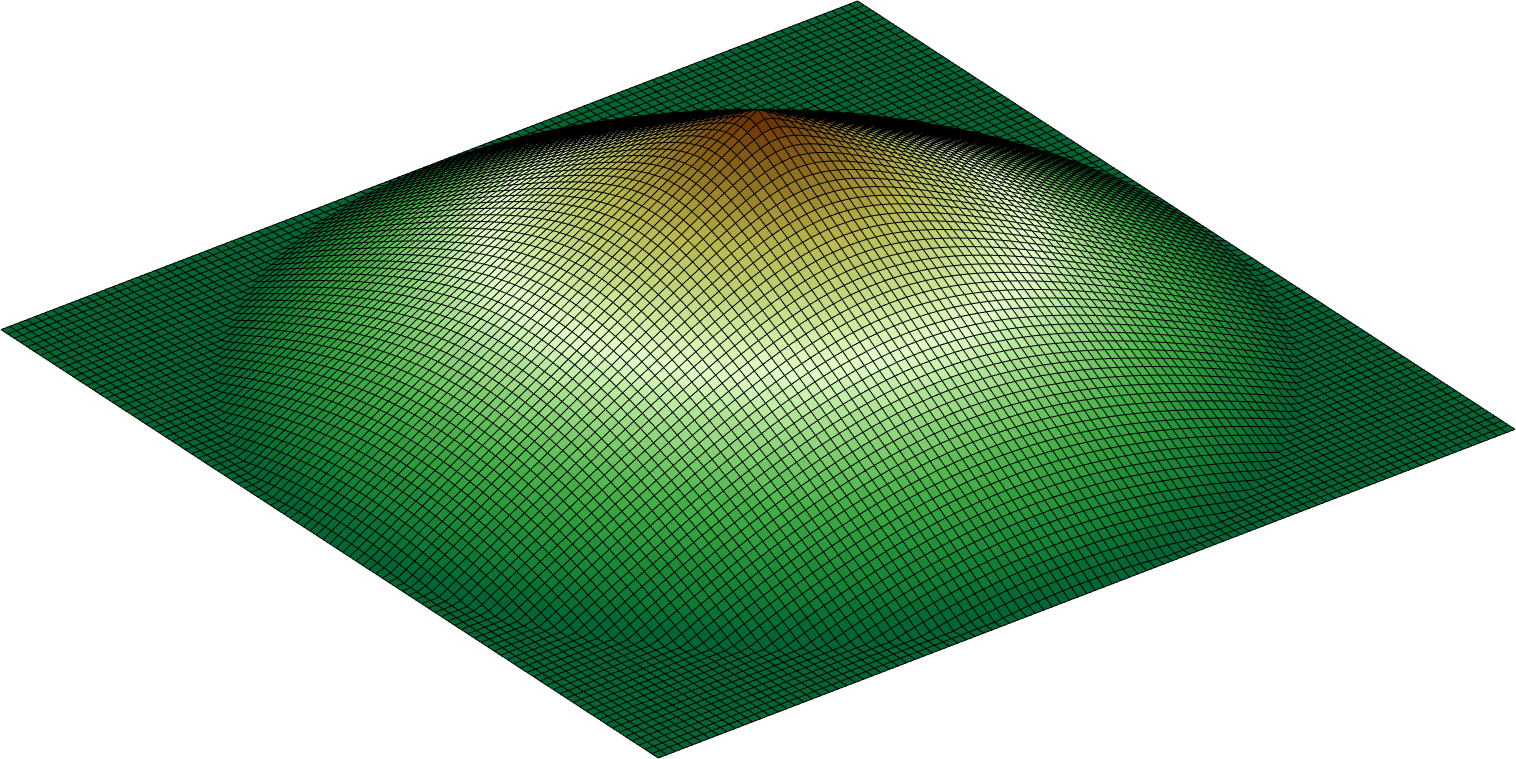}
    \includegraphics[width=0.32\columnwidth]{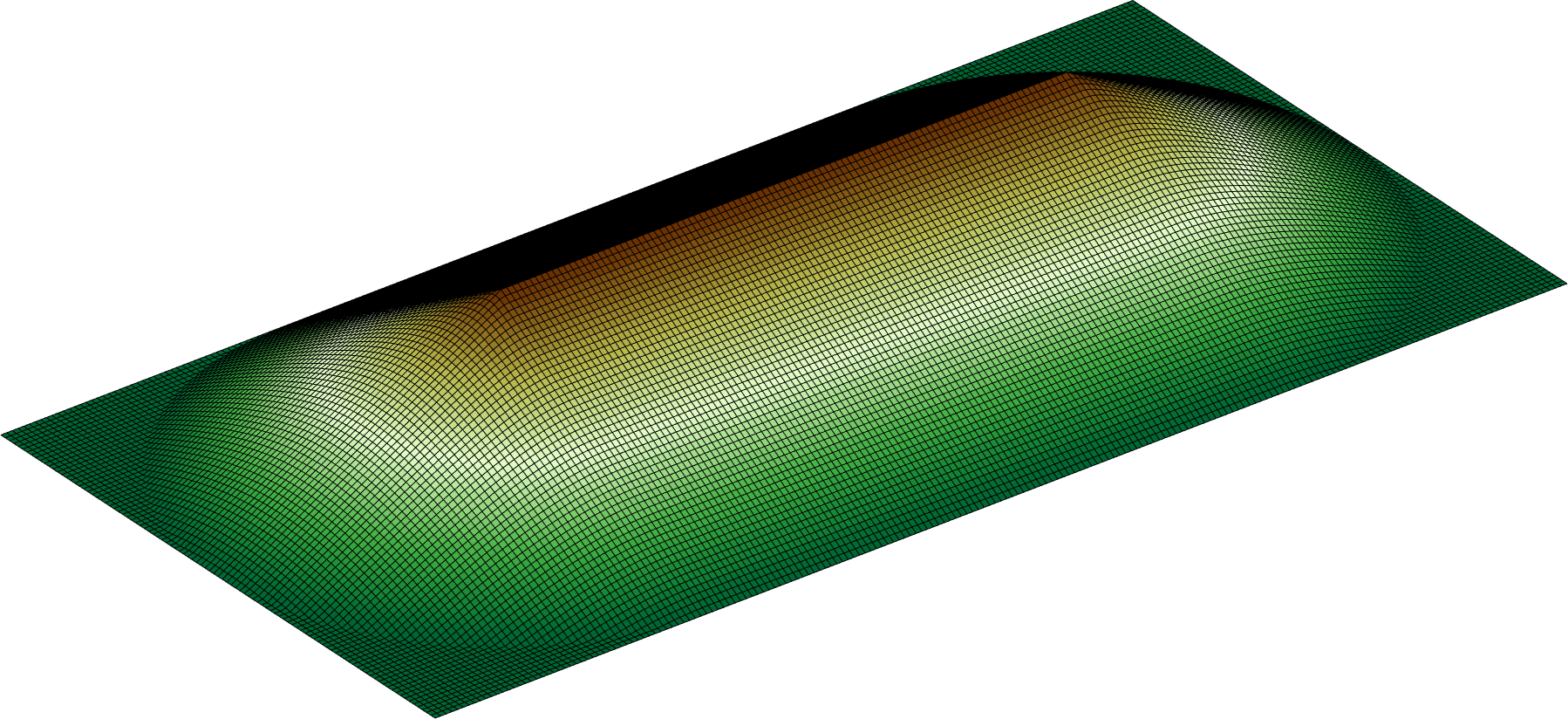}
    \includegraphics[width=0.32\columnwidth]{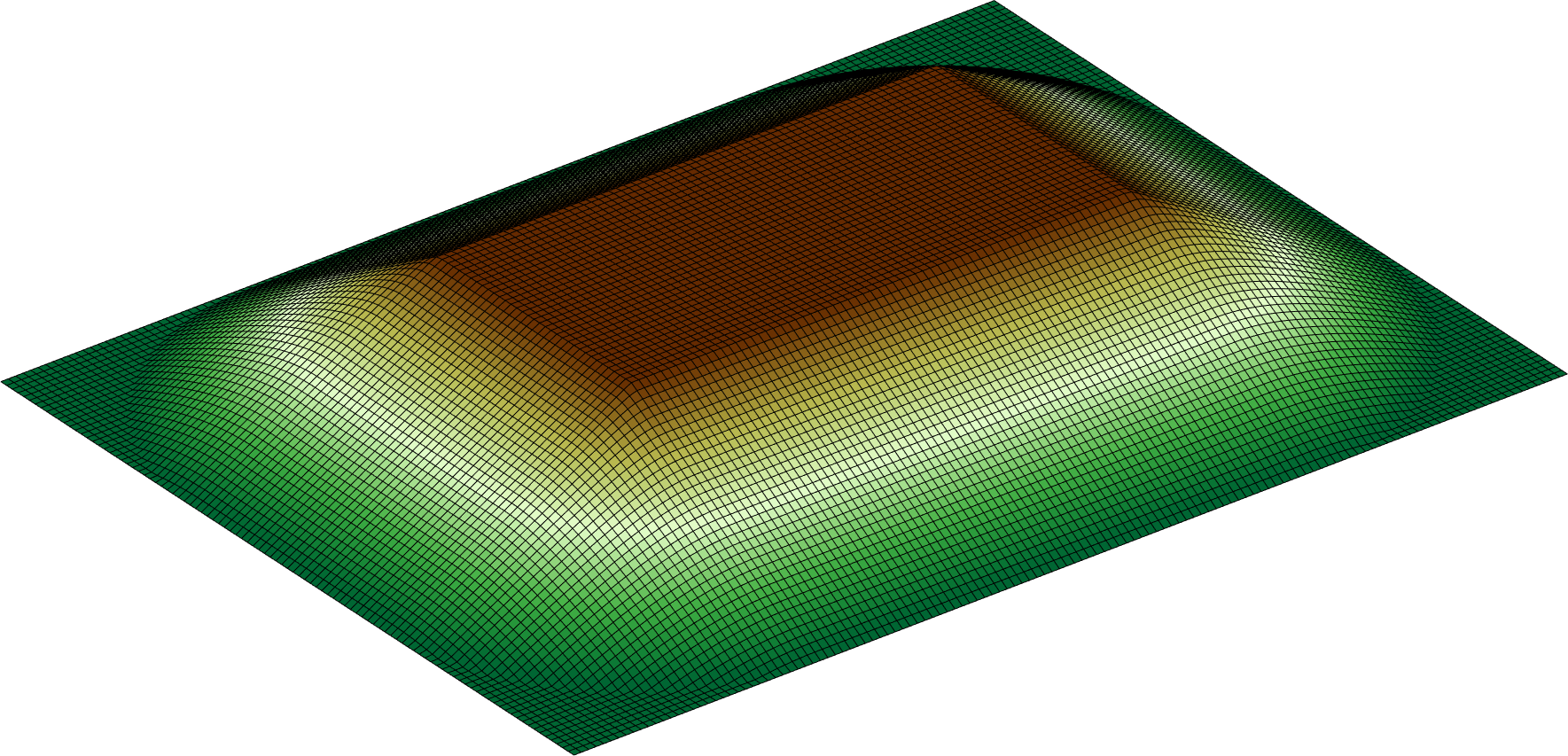}
    \caption{(Left to Right) Final form of the fixed height free base models when the granular material is released using a funnel-shaped unloader under 
    static 0D (point discharge), linear (along length dimension only) 1D and rectilinear 2D (along length and breadth dimensions) motion.}
\label{fig:3D_fixed-height}
\end{figure}

When the funnel-shaped unloader is stationary, the dry bulk material will form a right circular cone with a peak just below the unloader. For piles formed under motion of the unloader, i.e., it unloads, shifts, unloads, shifts, and so on, a grid pattern is imagined with the plane of motion containing the peaks of several overlapping
right-circular cones of the same height as shown in Fig.~\ref{fig:grid-patterns}. In the limit, the grid separation distance tends to zero, and we obtain the final 3D-shape of the pile as illustrated in Fig.~\ref{fig:3D_fixed-height}. \emph{The grid dimension classifies the pile-type.}

\begin{figure}[h!bt]
     \centering
     \begin{subfigure}[b]{0.3\columnwidth}
         \centering        \includegraphics[width=\textwidth]{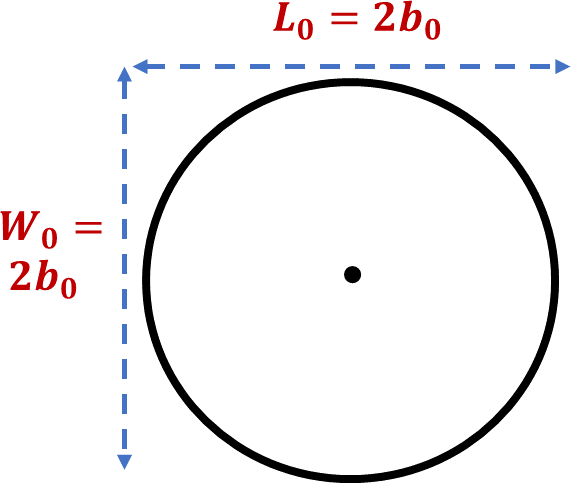}
        \caption{}\label{fig:fixed_0D}
     \end{subfigure}
     \hfill
     \begin{subfigure}[b]{0.5\columnwidth}
         \centering         \includegraphics[width=\textwidth]{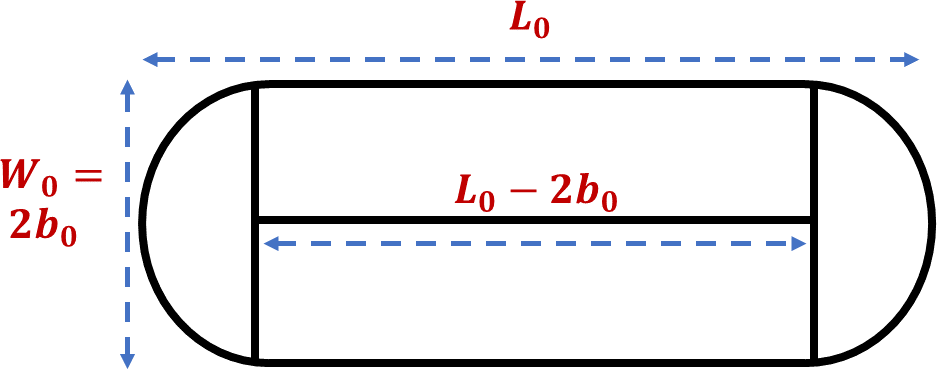}
         \caption{}\label{fig:fixed_1D}
     \end{subfigure}
     \hfill
     \begin{subfigure}[b]{0.5\columnwidth}
         \centering         \includegraphics[width=\textwidth]{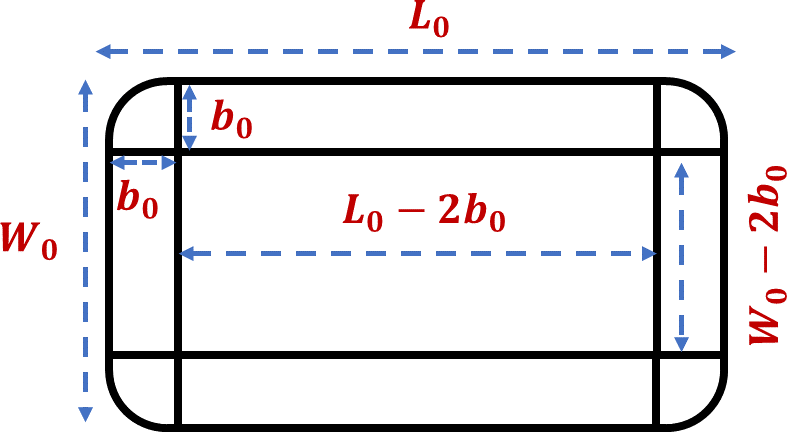}
         \caption{}\label{fig:fixed_2D}
     \end{subfigure}
        \caption{Footprint of fixed height free base models when the granular material is released using a funnel-shaped unloader under (\subref{fig:fixed_0D}) (point discharge), (\subref{fig:fixed_1D}) (along length dimension only) 1D and (\subref{fig:fixed_2D}) 2D (along length and breadth dimensions) motion.}
        \label{fig:fixed-height}
\end{figure}

We are interested in the extent of the footprint of the pile as shown in Fig.~\ref{fig:fixed-height}, which will be enough to extract its volume provided we know $H$ directly or $\theta_c$ corresponding to the material, from which we can calculate $H$ indirectly using~\eqref{eq:3par_reln}. 

Based on the footprints shown in Fig.~\ref{fig:fixed-height} we can estimate the volume ($V$) of 2D-pile in the following manner:
\begin{align}\label{2D_motion_model}
V  = &\int_{Height} \underbrace{({W_0 - 2b_0 \frac{x}{H}})({L_0 - 2b_0 \frac{x}{H}})}_{\substack{\text{Bounding Rectangle Area at height}\;x}}\;dx\\
    & -\int_{Height} \underbrace{({4-\pi})({b_0 - b_0 \frac{x}{H}})^2}_{\substack{\text{Area not part of the Pile at height}\;x}}\;dx\\
    = & \left(W_0L_0 - b_0(W_0 + L_0) + \frac{\pi}{3} b_0^2\right)H \label{Eq:Formula}
\end{align}
The above formulation implicitly assumes that ${L_0},{W_0} \geq b_0$. In addition, it results in a pile that has a rectangular top. For the 1D pile, $V$ is obtained by substituting $W_0 = 2b_0$ in~\eqref{2D_motion_model}, while for the 0D pile we substitute $L_0 = W_0 = 2b_0$.

When instead of $H$, $\theta_c$ of the material is known, with a simple substitution in~\eqref{Eq:Formula} using~\eqref{eq:3par_reln} we obtain the following working formula  
\begin{align}
     V = \left(b_0W_0L_0 - b_0^2(W_0 + L_0) + \frac{\pi}{3} b_0^3\right)\times\tan{\theta_c}\label{eq:Working_Formula}
\end{align}

\subsection{Practical Recipe}

Here we present a recipe on what practical steps are to be followed to estimate the volume of the piles with the proposed method from nadir-looking optical images (i.e., images from static camera set-up, imaging by flying drones or satellite imagery, etc.):
\begin{enumerate}
\item Detect piles in the image.
\item Determine if it is a regular pile of a fixed height.
\item Extract the footprint to determine $L_0, W_0, b_0$ of the pile.
\item Finally, \emph{estimate Volume}\begin{itemize} 
\item If $H$ is known or can be estimated from the image(s), calculate $V$ using~\eqref{Eq:Formula}.
\item \underline{Otherwise}, identify the material the pile is made up of. Use \emph{a priori} knowledge of critical angle of repose $\theta_c$ for the material of the pile.
And, calculate the $V$ using~\eqref{eq:Working_Formula}.
\end{itemize}
\end{enumerate}

\section{Demonstration on remote sensing image}
\begin{figure}[h!bt]
    \centering
    \includegraphics[width=0.48\columnwidth]{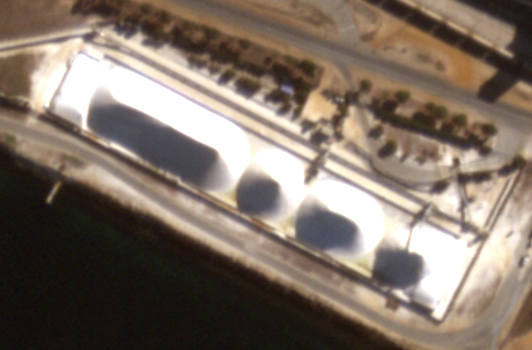}
    \includegraphics[width=0.48\columnwidth]{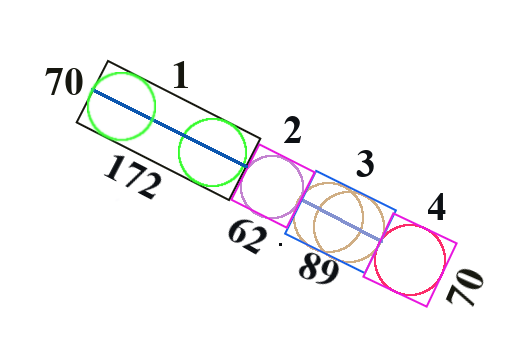}
    \caption{(L to R): Planet's SkySat imagery of \SI{50}{\cm} spatial resolution taken on April 19, 2024 from AustSand's silica sand storage facility ($35^\circ 02' 09'' \, \text{S}, \, 117^\circ 53' 58'' \, \text{E}$) on Princess Royal Drive at the Albany Port, Australia and a separate layer showing measured pile footprint dimensions expressed in pixel equivalent units.}
    \label{fig:silica_sand_piles}
\end{figure}

In order to demonstrate the method, we have acquired a Planet SkySat image of piles of silica sand, shown in  Fig~\ref{fig:silica_sand_piles}. The image is cropped from a larger RGB image, which is ortho-rectified, color-corrected, and super-resolved using a panchromatic band. 

As per publicly available information about port operations of AustSand Mining, it is mentioned that ``\emph{Approx 40 000 tonnes can be stored on site being made up of, up to four separate grades of silica sand.}"~\cite{AustSand_PO}. Our estimates from the image can be seen in Tab.~\ref{tab:Vol_Sand_Silica_Piles}.

\begin{table}[htb]
\centering
\caption{Estimating Weight of silica sand piles $\theta_c = 33.8^\circ$~\cite{anwar2021comparative} and bulk weight density = 1.6 \unit{T/m^3}~\cite{euroquarz2013silica}}
\label{tab:Vol_Sand_Silica_Piles}
\begin{tabular}{ccccccccc}
    \toprule
     Pile. no. & \multicolumn{2}{c}{$L_0$} & \multicolumn{2}{c}{$W_0$} & \multicolumn{2}{c}{$b_0$} & Weight\\
     & px & \unit{m} & px & \unit{m} & px & \unit{m} & \unit{kT}\\
     \midrule
     1 &  172 & 86 & 70 & 35 & 35 & 17.5 & 22.74 ($\approx$ 23)\\
     2 &  62 & 31 & 62 & 31 & 31 & 15.5 & 4.176 ($\approx$ 4)\\
    3 &  89 & 44.5 & 70 & 35 & 35 & 17.5 & 9.127 ($\approx$ 9)\\
     4 &  70 & 35 & 70 & 35 & 35 & 17.5 & 6.011 ($\approx$ 6)\\
     \bottomrule
\end{tabular}
\end{table}
Assuming a single bulk weight density, we get a total tonnage of $(23+4+9+6)=42$ in approximate kilotonnes (\unit{kT}) ($= 10^9$ \unit{g}) which is within 5 \% error with the maximum capacity of stored silica sand piles (40 \unit{kT}) as determined by their port operation regulations. The estimation could be more accurate if bulk density of the four grades of silica sand is known \emph{a priori} as well as the grade to which each pile corresponds.

\section{Discussion}

In the analysis presented above, we have not investigated how the border or footprint of the pile should be determined. For remote sensing satellite images, one of the important issues that will arise in the real world is the location of any shadows, obstruction of view due to presence of other massive objects, and what material the ore sits on. Another will be the resolution needed to achieve good boundary detection, as a lower resolution will mean that the curvature and lengths will be harder to determine. The combination of resolution and pile size will form an interplay in determining the errors in boundary extraction and measurement. To overcome some of these challenges, one way is to train a machine learning segmentation algorithm on the types of ore of interest. Once these boundaries have been determined, our presented closed-form analytic method will be faster than techniques that try to reconstruct the 3D shape of the piles requiring many point measurements/estimates, in turn increasing the memory needs and computational time. 

\section{Conclusion}
In this work, we have demonstrated the effectiveness of our proposed volume estimation model from a single satellite image in the context of dry bulk cargo pile to good accuracy. The achieved accuracy in total dry bulk cargo storage is within $95\%$. It also avoids reconstructing the 3D shape from image, making it simple and direct provided pile footprint dimensions are known or can be estimated from the imagery. The proposed method in general, works for satellite imagery and also for downward looking static camera views and imagery by flying drones subject to correction for geometric distortions that may arise during image formation. In future, the proposed methodology will be combined with machine learning techniques to study irregular and reclaimed piles of different ore materials for segmenting and estimating their footprint dimensions towards  estimation of dry bulk cargo pile volumes from a single optical image.

\section*{Acknowledgment}
This work is supported by the NFR-funded SHIPTRACK  project (Project code: 326609).

\bibliographystyle{IEEEtran}
\bibliography{mybibfile}  

\begin{thebibliography}{1}
\providecommand{\url}[1]{#1}
\csname url@samestyle\endcsname
\providecommand{\newblock}{\relax}
\providecommand{\bibinfo}[2]{#2}
\providecommand{\BIBentrySTDinterwordspacing}{\spaceskip=0pt\relax}
\providecommand{\BIBentryALTinterwordstretchfactor}{4}
\providecommand{\BIBentryALTinterwordspacing}{\spaceskip=\fontdimen2\font plus
\BIBentryALTinterwordstretchfactor\fontdimen3\font minus
  \fontdimen4\font\relax}
\providecommand{\BIBforeignlanguage}[2]{{%
\expandafter\ifx\csname l@#1\endcsname\relax
\typeout{** WARNING: IEEEtran.bst: No hyphenation pattern has been}%
\typeout{** loaded for the language `#1'. Using the pattern for}%
\typeout{** the default language instead.}%
\else
\language=\csname l@#1\endcsname
\fi
#2}}
\providecommand{\BIBdecl}{\relax}
\BIBdecl

\bibitem{alsayed2023stockpile}
A.~Alsayed and M.~R. Nabawy, ``Stockpile volume estimation in open and confined
  environments: A review,'' \emph{Drones}, vol.~7, no.~8, p. 537, 2023.

\bibitem{dAutume2020Stockpile}
M.~d’Autume, A.~Perry, J.-M. Morel, E.~Meinhardt-Llopis, and G.~Facciolo,
  ``Stockpile monitoring using linear shape-from-shading on planetscope
  imagery,'' \emph{ISPRS Annals of the Photogrammetry, Remote Sensing and
  Spatial Information Sciences}, vol.~2, pp. 427--434, 2020.

\bibitem{Marchesconi2022Interactive}
F.~Marchesoni-Acland, M.~d'Autume, G.~Facciolo, C.~De~Franchis, J.-M. Morel,
  and E.~Meinhardt-Llopis, ``Interactive segmentation for shape from shading
  over hr sar images,'' in \emph{IGARSS 2022 - 2022 IEEE International
  Geoscience and Remote Sensing Symposium}, 2022, pp. 975--978.

\bibitem{gitau2022spatial}
F.~Gitau, J.~K. Maghanga, and M.~N. Ondiaka, ``Spatial mapping of the extents
  and volumes of solid mine waste at samrudha resources mine, kenya: a gis and
  remote sensing approach,'' \emph{Modeling Earth Systems and Environment},
  vol.~8, no.~2, pp. 1851--1862, 2022.

\bibitem{Mari2021Automatic}
R.~Marí, C.~De~Franchis, E.~Meinhardt-Llopis, and G.~Facciolo, ``Automatic
  stockpile volume monitoring using multi-view stereo from skysat imagery,'' in
  \emph{2021 IEEE International Geoscience and Remote Sensing Symposium
  IGARSS}, 2021, pp. 4384--4387.

\bibitem{kappaupsilon1966notes}
H.~H. Ku, ``Notes on the use of propagation of error formulas,'' \emph{Journal
  of Research of the National Bureau of Standards: Engineering and
  instrumentation. Section C.}, vol.~70, no.~4, p. 263, 1966.

\bibitem{AustSand_PO}
\BIBentryALTinterwordspacing
``{AustSand Mining Port Operations},'' 2025, [Accessed: 2025-03-17]. [Online].
  Available: \url{https://www.austsandmining.com.au/port-operations/}
\BIBentrySTDinterwordspacing

\bibitem{anwar2021comparative}
N.~Anwar, T.~Sappinen, K.~Jalava, and J.~Orkas, ``Comparative experimental
  study of sand and binder for flowability and casting mold quality,''
  \emph{Advanced Powder Technology}, vol.~32, no.~6, pp. 1902--1910, 2021.

\bibitem{euroquarz2013silica}
\BIBentryALTinterwordspacing
``{Silica Sand Dry} data sheet overview,'' 2024, [Accessed: 2024-12-15].
  [Online]. Available:
  \url{https://www.euroquarz.com/uploads/tx_datasheetoverview/Silica_sand_dry_0_5-1_2_mm__plant_11__EN.pdf}
\BIBentrySTDinterwordspacing

\end{thebibliography}
\end{document}